\documentclass[prl,aps,twocolumn,floatfix,longbibliography]{revtex4-2}

\usepackage{amssymb,mathrsfs,amsfonts}
\usepackage{amsmath}
\usepackage{epstopdf}
\usepackage{listings}
\usepackage{textcomp}
\usepackage{bibnames}
\usepackage{appendix}
\usepackage{epstopdf}
\usepackage{nicefrac}
\usepackage{listings}
\usepackage{upgreek}
\usepackage{physics}
\usepackage{multirow}
\usepackage{color}
\usepackage{natbib}
\usepackage{graphicx}
\usepackage{tabls}
\usepackage{afterpage}
\usepackage{amsthm}
\usepackage{lastpage}
\usepackage{empheq, etoolbox}
\usepackage{epstopdf}
\usepackage{changes} %
\colorlet{Changes@Color}{red}  

\usepackage{orcidlink}

\begin{document}

\title
{An accurate reaction-diffusion limit to the spherical-symmetric Boltzmann equation}
\author
{Shay I. Heizler\orcidlink{0000-0002-9334-5993}}
\thanks{These authors contributed equally to this work.}
\author
{Menahem Krief}
\thanks{These authors contributed equally to this work.}
\author
{Michael Assaf}

\affiliation{Racah Institute of Physics, The Hebrew University, 9190401 Jerusalem, Israel}

\begin{abstract}

We resolve a long standing question regarding the suitable effective diffusion coefficient of the spherically-symmetric transport equation, which is valid at long times. To that end, we generalize a transport solution in three dimensions for homogeneous media, to include general collisional properties, including birth-death events and linearly anisotropic scattering. This is done by introducing an exact scaling law relating the Green function of the pure-scattering case with the general collision case, which is verified using deterministic and Monte-Carlo simulations. Importantly, the effective diffusion coefficient is identified by inspecting the transport solution at long times.

\end{abstract}

\maketitle

\textit{Introduction.}
The study of three-dimensional (3D) time-dependent transport phenomena is at the heart of many fields in physics, chemistry and biology. For example in astrophysics, when a massive star explodes, there is a blast of energy that is released in a relatively focused spot~\cite{castor2004,levpom}. Similarly, in inertial confinement fusion a large focused release of energy occurs as a result of fusion reactions~\cite{ICF,ICF2,lindl}. In these cases, X-ray photons are propagating fast toward the medium due to the radiative transfer mechanism. Correctly modeling the propagation of these heat waves is a prerequisite for analyzing the outgoing signals, e.g., from a supernovae explosion. Time-dependent transport phenomena also appear in electron transport in hot plasmas~\cite{Spitzer}, modeling coda waves of local earthquakes~\cite{Carcole}, the distribution of neutrons inside a nuclear reactor~\cite{CaseZweifel1967,ganapol1985,DullaRavetto}, and optical properties of biological tissues in biology~\cite{Hulst1996,Durduran1997,Corngold,Graaff2000}. Notably, in the latter, the balance between scattering and absorption inside the medium are of great interest~\cite{Hoenderes}. Even in modeling the transport of bacteria that diffuse in a medium and may reproduce or die, a transport equation can be defined to model the population dynamics~\cite{othmer2000diffusion}.


The exact particle propagation is modeled via the Boltzmann (transport) equation for the probability density function (PDF) $P(\vec{r}, t, \hat{\Omega})$ per unit volume $d^{3}\vec{r}$ and direction $d\hat{\Omega}$, at time $t$. Notably, in most of the fields specified above, the governing equation of transport is of non-interacting particles, which are aptly described by the {\em{linear}} Boltzmann equation.
When the medium is highly scattering, the exact solution of this equation includes a diffusive region in the bulk, and ballistic `tails' due to particles that almost do not undergo collisions~\cite{METZLER20001,pesach,stas}. However, obtaining the solution of the transport equation in the general case is not amenable analytically, and numerically it is highly time consuming. Notably, in a one-dimensional (1D) geometry, where the particles are free to travel only forward and backward, the linear transport equation reproduces the Telegrapher's equation, which can be solved analytically~\cite{goldstein,weiss,masoliver,pesach,METZLER20001}. 

It is well known that at sufficiently long times, the exact transport solution tends to a reaction-diffusion solution via the central limit theorem, due to multiple scattering events. The central part of the distribution function has a Gaussian shape with increasing width, which corresponds to an effective diffusion coefficient. Naturally, applying a diffusion approximation to the transport equation gives rise to non-physical tails of particles at $r\!>\!vt$~\cite{Heizler2010}. Yet, even in the bulk region where the diffusion approximation holds, it remains an open question of which diffusion coefficient to use to accurately reproduce the bulk solution of the full transport equation.

In some limits, e.g., in the purely scattering case, the effective diffusion coefficient is well known.  
The classic definition of the diffusion approximation uses the total mean free path (mfp), including absorption and scattering events~\cite{case1953,CaseZweifel1967,BellGlasstone,Durduran1997,Corngold}. On the other hand, some studies show that the diffusion coefficient should only depend on the scattering mfp~\cite{Furutsu2,levpom,pomraning1981,ganapol1985,Furutsu1,Tsuchiya1995,Durduran1997,zoia}. When birth events are also included, a modified diffusion coefficient, accurate in the limit of source-dominant media, was offered in modeling the radiative transfer in astrophysics~\cite{levpom}.
Yet, for an arbitrary collision scenario, or for highly anisotropic scattering, the corresponding effective diffusion coefficient is unknown~\cite{levpom,pomraning1981,Furutsu2,ganapol1985,Furutsu1,Tsuchiya1995,Durduran1997,Graaff2000,Corngold,Hoenderes2,Hoenderes1,DullaRavetto,zoia,Liemert,Liemert2}.

Here, we present a simple yet rigorous derivation of the accurate diffusion coefficient, valid in the spherical symmetric case, in the presence of an arbitrary set of processes, including birth (or branching) and death (or absorption) events, that may occur upon an interaction between an agent and the medium. 
This diffusion limit is crucial, e.g., in modeling  radiative transfer in the non-LTE (Local Thermodynamic Equilibrium) regime, when photon emission dominates absorption~\cite{levpom}, which occurs in various astrophysical scenarios, such as shock breakout phase during supernovae, corona of accretion disks and nebular phase of stellar explosions, to name a few.

Our analysis is based on the generalization of the solution of~\citet{pesach}, which was obtained for homogeneous infinite medium and pure isotropic scattering. While in~\cite{Martelli} the solution was extended to the case of linear anisotropic scattering, we here generalize the solution to hold for an arbitrary collision scenario. We first derive an exact scaling relation (in space and time) of the 3D transport solution between the pure-scattering and general-collision case, which holds for the general anisotropic case as well, and find the Green function solution for a point source term. The scaling relation is verified via two different numerical schemes: (i) a solution of the deterministic time-dependent equation with the discrete ordinates method ($S_N$ method)~\cite{lewismiller}, and (ii) stochastic Monte-Carlo (MC) simulations mimicking the different probabilistic events, see supplemental material (SM), Sec.~I. This scaling relation allows us to extend the analytical solution of Paasschens for the general-collision case, from which we infer the  diffusion coefficient by computing its long time asymptotics.

\textit{Spherical Symmetric Transport Solution.}
In the case of no external forces and assuming mono-energetic particles, the equation that governs the physical process is the {\em{linear}} Boltzmann transport equation. For a spherical-symmetric setting, the transport equation takes the following form~\cite{CaseZweifel1967,BellGlasstone,lewismiller,pesach} (see SM, Sec.~II):
\begin{align} \label{spherical}
    &\left(\frac{\partial}{v\partial t} + \mu\frac{\partial}{\partial r} +\frac{1-\mu^2}{r}\frac{\partial}{\partial\mu} + \ell_t^{-1}\right)P(r, t, \mu) = \\
    &\frac{c(r,t)\ell_t^{-1}}{2}\int_{-1}^1 d\mu f(\mu_0)P(r, t, \mu) + Q_{\mathrm{ext}}(r, t, \mu)/v. \nonumber
\end{align}
Here the PDF, $P(r, t, \mu)$, is a function of the radius $r$, time $t$, and the direction of the particle with respect to $r$, $\mu\equiv\hat{\Omega}\cdot\hat{r}$,  where $\hat{\Omega}$ is the direction of the particle before the collision. In addition, we have assumed that the interaction between particles is negligible, and the medium is locally in equilibrium, such that the Boltzmann collision term can be modeled using the definitions of the mfps for different interactions (or alternatively cross sections). These are the mfp of absorption, $\ell_a$, and of scattering, $\ell_s$, where we denote by $\ell_t$ the total mfp due to all physical events. In general, the scattering event is a function of the angular deflection during scattering, i.e., the angle between the direction of the particle before the scattering $\hat{\Omega}$ and after $\hat{\Omega}'$~\footnote{Note that in this spherical-symmetric case, the particle changes its direction $\mu$ even by undergoing a pure ballistic motion from $\vec{r}_1$ to $\vec{r}_2$.}. As a result, the \textit{macroscopic cross section} for scattering has the form $\Sigma_{s}(\hat{\Omega}\to\hat{\Omega'})=\ell_s^{-1}f(\hat{\Omega}\cdot\hat{\Omega}')=\ell_s^{-1}f(\mu_0)$,  where $f(\mu_0)$ is a normalized distribution of the deflection angle cosine 
 $\mu_0\equiv\hat{\Omega}\cdot\hat{\Omega}'$. The function $f$ is problem dependent, and is commonly taken in the Henyey-Greenstein form~\cite{henyey} (see SM, Sec.~IVa)~\footnote{Here, the function $f$ is taken to be of the form:
$f(\mu_0)=(1/2)(1-g^2)/(1+g^2-2g \mu_0)^{3/2}$, where $g=\bar{\mu}_0$ denotes the average scattering angle cosine~\cite{Martelli}.}. Therefore, the collision term in Eq.~(\ref{spherical})
can be written explicitly as a ``gain-loss" term using the above mfps definitions. 

Additional notations in Eq.~(\ref{spherical}) include the particle velocity $v$ and an external source $Q_{\mathrm{ext}}(r, t, \mu)$, which may be a function of space, direction and time. 
Finally, $c(r,t)$ represents the mean number of particles that are emitted in an interaction, including sources~\cite{case1953,Kuscer1965,CaseZweifel1967,BellGlasstone,doyas,ganapol1985,Heizler2010,DullaRavetto,winslow,eugene}, for example due to cell division, and satisfies~\footnote{Here, $c$ is often called $\omega_{\mathrm{eff}}$ in radiative transfer~\cite{pomraning1981,levpom} and $a$ in optics~\cite{Hulst1996,Graaff2000,Hoenderes1,Hoenderes2,Hoenderes}).}:
\begin{equation}
    c(r,t) = \frac{\ell_s^{-1}{\cal P}(r,t) + \frac{1}{2}\int_{-1}^1 Q_{\mathrm{int}}(r, t, \mu)d\mu/v}{\ell_t^{-1} {\cal P}(r,t)}.
\label{omega}
\end{equation}
Here ${\cal P}(r,t)=\frac{1}{2}\int_{-1}^1 d\mu P(r, t, \mu)$ is the population probability density, which depends only on the magnitude of $r$ and time. In the case of no internal source and homogeneous media,  $0\leqslant c \leqslant 1$ becomes the ratio of the total and scattering mfps, $c=\ell_s^{-1}/\ell_t^{-1}=\ell_s^{-1}/(\ell_s^{-1}+\ell_a^{-1})$. If the medium involves internal sources such as birth events proportional to ${\cal P}(\vec{r},t)$, one has $c=(\ell_s^{-1}+\bar{n}\ell_b^{-1})/\ell_t^{-1}=(\ell_s^{-1}+\bar{n}\ell_b^{-1})/(\ell_s^{-1}+\ell_a^{-1}+\ell_b^{-1})$ which may exceed $1$. Here $\bar{n}$ is the mean number of particles created in a birth event and $\ell_b$ is the birth mfp. Clearly, in systems with $c<1$ ($c>1$) the expected number of particles decays (exponentially grows) in time. For example, in astrophysics, the emitted black-body energy density $B$ and  radiation energy density $E$ allow us to define $c$ as $c\equiv\omega_{\mathrm{eff}}=(\ell_a^{-1}B+\ell_s^{-1}E)/\ell_t^{-1}E$.
While in LTE, $E\approx B$ and thus $\omega_{\mathrm{eff}}\approx 1$, in non-LTE, when the matter is hotter than the radiation, $B>E$ and $\omega_{\mathrm{eff}}>1$, and vice versa~\cite{levpom}. Naturally, when $c>1$, the exponential growth is eventually arrested by nonlinear effects.

Henceforth, we will refer to the cases without and with internal sources as the sub-scattering and super-scattering problem, respectively. We will also focus on the case of infinite homogeneous medium; i.e., we assume $c(r,t)=c$ is a constant. Importantly, we perform our calculations below using an external point-like source at the origin: $Q_{\mathrm{ext}}(r,t)=\delta(r)\delta(t)/4\pi r^2$, where $\delta(x)$ is the Dirac delta function. In this case, the calculated population density will serve as the Green function, which will allow the subsequent calculation of the population density for any space- and time-dependent external source term by the Green's convolution integral.
 
In Fig.~\ref{sn_mc1} we compare our analytical solution to Eq.~(\ref{spherical}), see Eq.~(\ref{pesach_aniso}) below, to numerical solutions using the $S_N$ and MC methods. Here we show ${\cal P}(r,t)$ as a function of the rescaled position $\tilde{r}=r/\ell_t$, at different rescaled times $\tilde{t}=tv/\ell_t$, for various values of $c$, $0.5\leqslant c\leqslant 2$. Both the analytical and numerical solutions show ballistic behavior near $r=vt$. Yet, the ballistic region shrinks and the diffusive behavior becomes dominant as time advances.
\begin{figure}
\includegraphics[width=8.50cm,clip=true]{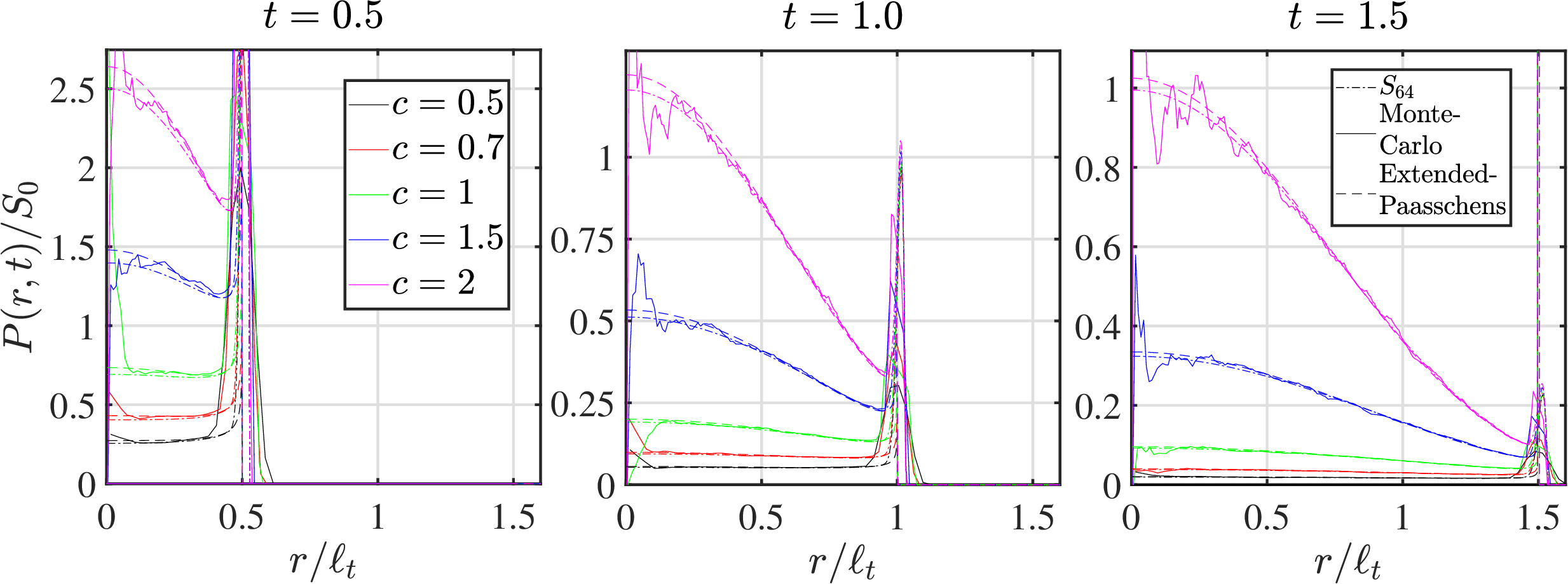}
\vspace{-4mm}\caption{The population density ${\cal P}(r,t)$ at different times for various values of $c$ [Eq.~(\ref{omega})], see legend, in the isotropic case. Our analytical solution, Eq.~(\ref{pesach_aniso}) (dashed lines) is compared with the numerical solution using the $S_N$ (dash-dotted lines) and MC (solid lines) methods. 
The noisy solution at $r\simeq 0$ is due to the small volumes of the numeric cells near the origin. }  
\label{sn_mc1}
\vspace{-5mm}
\end{figure}

\begin{figure}
\includegraphics[width=8.10cm]{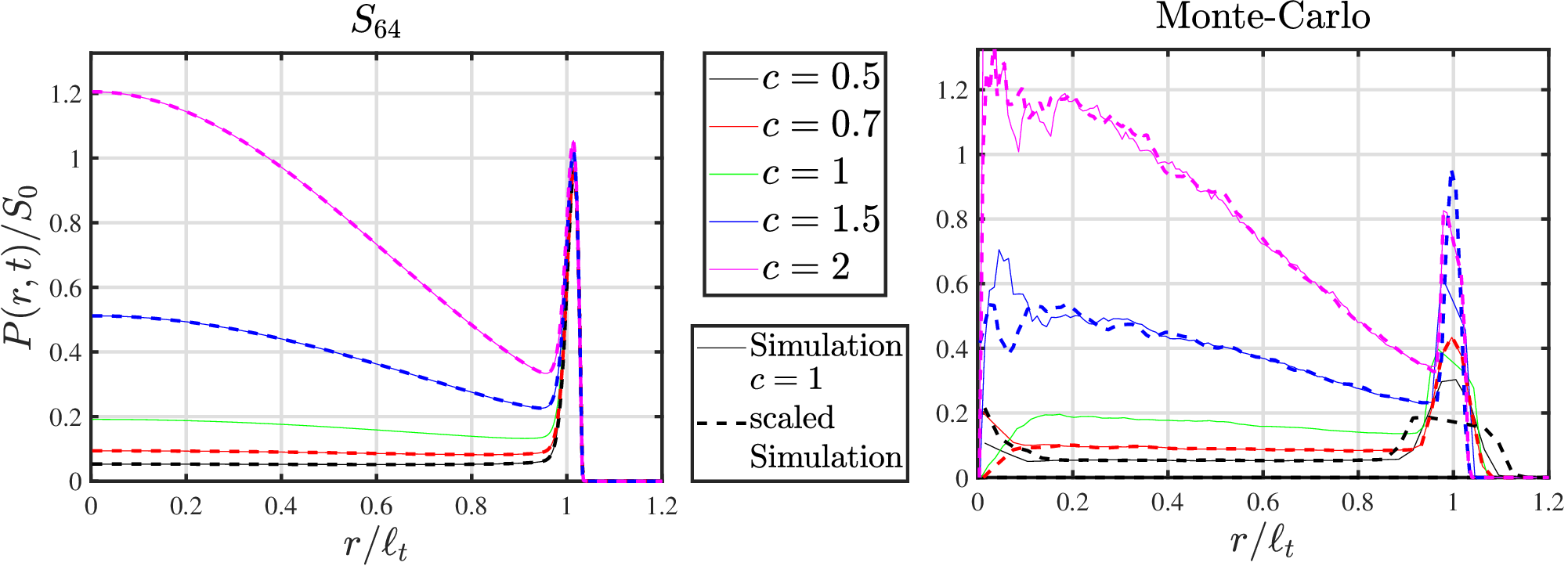}
\vspace{-3mm}\caption{A numerical verification of the scaling relation [Eq.~(\ref{scale2})] for various values of $c$. The solid lines represent the $S_N$ (left) and MC (right) results for the population densities, calculated with a specific value of $c$, while the dashed lines are the rescaled results with $c=1$, using Eq.~(\ref{scale2}).} 
\label{sn_mc2}
\vspace{-2mm}\end{figure}

We now solve Eq.~(\ref{spherical}) is the general collision scenario, i.e., $c\neq 1$. To do so, we identify a scaling relation between the PDF, $P(r,t,\mu; c)$, in the general $c\ne1$ case, and the PDF, $P(r,t,\mu; 1)$, in the pure-scattering, $c=1$, case. Previously, it was shown that for $c\leqslant1$, the scaling relation reads~\cite{Martelli,pesach,zoia}: $P(r,t,\mu;c)=e^{-vt/\ell_a}P(r,t,\mu;1)$, with
$c=\ell_s^{-1}/(\ell_a^{-1}+\ell_s^{-1})$.
We propose the following exact scaling relation for the Boltzmann equation in spherical symmetry~(\ref{spherical}),
for a general value of $c$ (see SM, Sec.~III):
\begin{equation}
P(r,t,\mu;c)=c^3e^{-(1-c)vt/\ell_t}P(cr,ct,\mu;1)
\label{scale2}
\end{equation}
where $\ell_t$ denotes the total mfp. In Fig.~\ref{sn_mc2} we numerically verify the scaling relation by plotting ${\cal P}(r,t)$ using the $S_N$ and MC methods, for various values of $c$, both in the ballistic and diffusion regions.

We now use this scaling relation to find the solution for the population density ${\cal P}(r,t; 1)$, for $c\ne1$. Here, our starting point is the result in the pure-scattering case, $c=1$, obtained by~\citet{pesach}. This result assumes a homogeneous infinite medium, and was obtained exactly for 2D and 4D geometries using a Fourier transforms in space and time. Interestingly, this method is less useful in the 3D case as the inverse transform cannot be found analytically. Instead, Paasschens used an interpolation between the 2D and 4D solutions~\cite{pesach}, which yields ${\cal P}(r,t;1)$ in the 3D case:
\begin{eqnarray}\label{pes_sol}
{\cal P}(r,t;1)&\simeq&\frac{e^{-vt/\ell_s}}{4\pi r^2}\delta(r-vt)+\frac{(1-r^2/v^2t^2)^{\nicefrac{1}{8}}}{(4\pi\ell_s vt/3)^{\nicefrac{3}{2}}}e^{-vt/\ell_s}\nonumber\\
&\times&G\left(\frac{vt}{\ell_s}\left[1-\frac{r^2}{v^2t^2}\right]^{\nicefrac{3}{4}}\right) \Theta(vt-r)\nonumber\\
G(x)&=&8(3x)^{-\nicefrac{3}{2}}\sum_{N=1}^{\infty}\frac{\Gamma\left(\frac{3}{4}N+\frac{3}{2}\right)}{\Gamma\left(\frac{3}{4}N\right)}\frac{x^N}{N!}
,
\end{eqnarray}
where $\Theta(r)$ is the Heaviside step function and $\Gamma(x)$ is the gamma function. Notably, see below, at long times Eq.~(\ref{pes_sol}) converges to the diffusion solution with $D\!=\!v\ell_t/3$ and $c=1$. Solution~(\ref{pes_sol}) can be extended to the linear anisotropic scattering case, upon using $\ell_{\mathrm{tr}}$ instead of $\ell_t$, where $\ell_{\mathrm{tr}}^{-1}\equiv\ell_t^{-1}(1-g)$, and $g=\bar{\mu}_0$ denotes the average scattering angle cosine~\cite{Martelli}.
Here,  $\ell_{\mathrm{tr}}$ is called the {\em{transport mfp}}~\cite{BellGlasstone}. However, we show that this solution is valid only at long times (see SM Sec.~IVa). 

With the scaling relation~(\ref{scale2}) we have found, and using Eq.~(\ref{pes_sol}) for $c=1$, the general solution for the population density, ${\cal P}(r,t;c)$ for $c\neq 1$ reads (see SM, Sec.~IVb):
\begin{align}
\label{pesach_aniso}
&{\cal P}(r,t;c)\simeq\frac{e^{-vt(1\!-\!cg)/\ell_t}}{4\pi r^2}\delta(r\!-\!vt)+
\frac{(1-r^2/v^2t^2)^{\nicefrac{1}{8}}}{(4\pi\ell_tvt/[3c(1\!-\!g))]^{\nicefrac{3}{2}}} \\ &\times G\left(\frac{vct(1-g)}{\ell_t}\left[1-\frac{r^2}{v^2t^2}\right]^{\nicefrac{3}{4}}\right) e^{-vt(1-cg)/\ell_t} \Theta(vt-r).\nonumber
\end{align}
This solution coincides with the Paasschens solution for $c=1$ and $g=0$. It can be shown to be valid for $g\ll1$ at all times; yet, it becomes valid for all values of $g$, $-1<g<1$, at long times (see SM, Sec.~IV).

\begin{figure}
\includegraphics[width=8.5cm,clip=true]{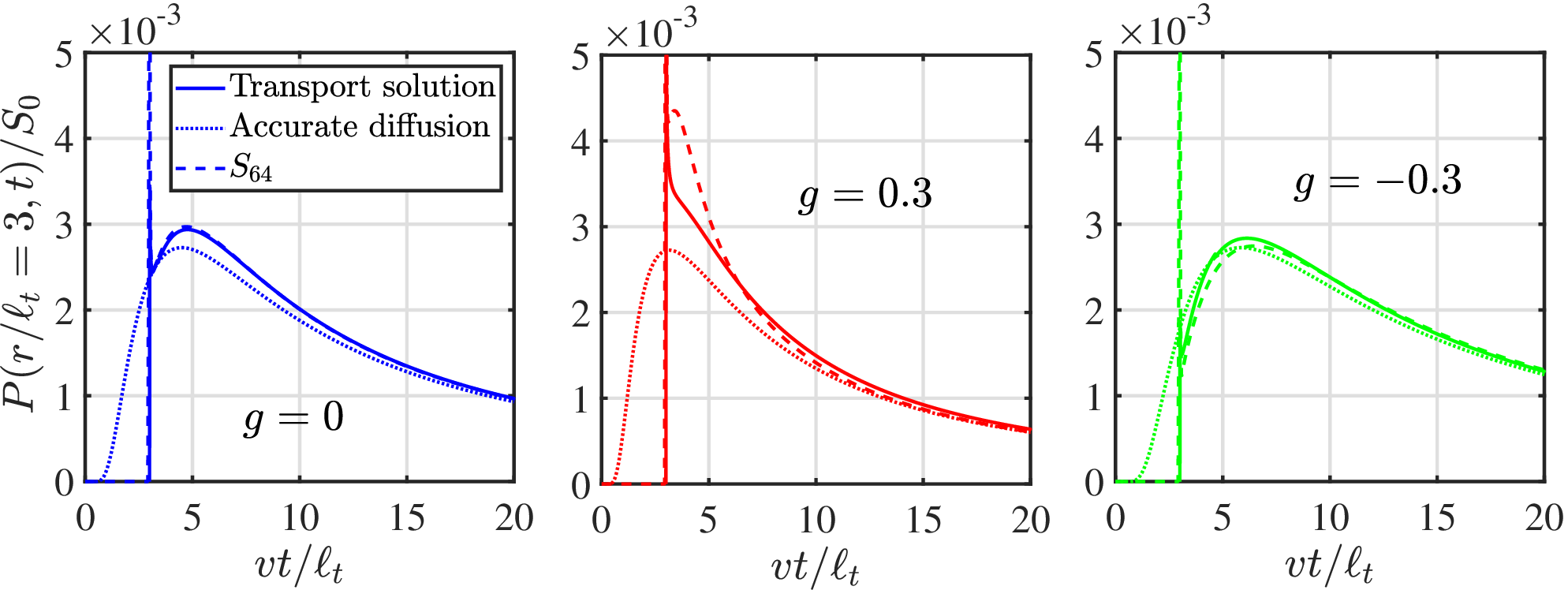}
\vspace{-3mm}\caption{The population density ${\cal P}(r,t)$ versus the normalized time $vt/\ell$, at $\tilde{r}=3$ (right), for various values of anisotropy $g$. Shown are the results obtained from the extended Paasschens solution (solid lines), $S_N$ simulations (dashed lines) and the accurate diffusion~\cite{ganapol1985} (dotted lines).}
\label{aniso}
\end{figure}

In Fig.~\ref{sn_mc1} our analytical solution [Eq.~(\ref{pesach_aniso})]  is shown to excellently agree with the numerical results of both $S_N$ and MC, for all values of $c$, in the isotropic case. In Fig.~\ref{aniso} we test the accuracy of Eq.~(\ref{pesach_aniso}) for the general anisotropic case for $c=1$. Here, we plot the PDF for the dimensionless spatial position $\tilde{r}=3$, as a function of $\tilde{t}$ for isotropic ($g=0$), forward  ($g=0.3$) and backward ($g=-0.3$) scattering. Importantly, we have checked that our results exactly reproduce earlier benchmark results for these $g$ values~\cite{ganapol1985}. While for $g=0$ our solution coincides with Eq.~(\ref{pes_sol}), and is thus accurate at all times, for $g\neq 0$ the picture is different. For forward scattering ($g=0.3$) Eq.~(\ref{pesach_aniso}) yields a good agreement only at late times, whereas for  backward scattering ($g=-0.3$), Eq.~(\ref{pesach_aniso}) yields a good agreement also for earlier times, see Fig.~\ref{aniso}. This is because forward scattering increases anisotropy, thereby delaying the validity of the diffusion assumption. Conversely, backward scattering causes particles to return to the origin, thereby increasing isotropy. It is also demonstrated that for the various values of $g$, the diffusion solution yields solutions that propagate faster than the particle speed at early times, and converge to the exact solution at long times.

\begin{figure}[t]
\begin{center} 
\includegraphics[width=8.8cm]{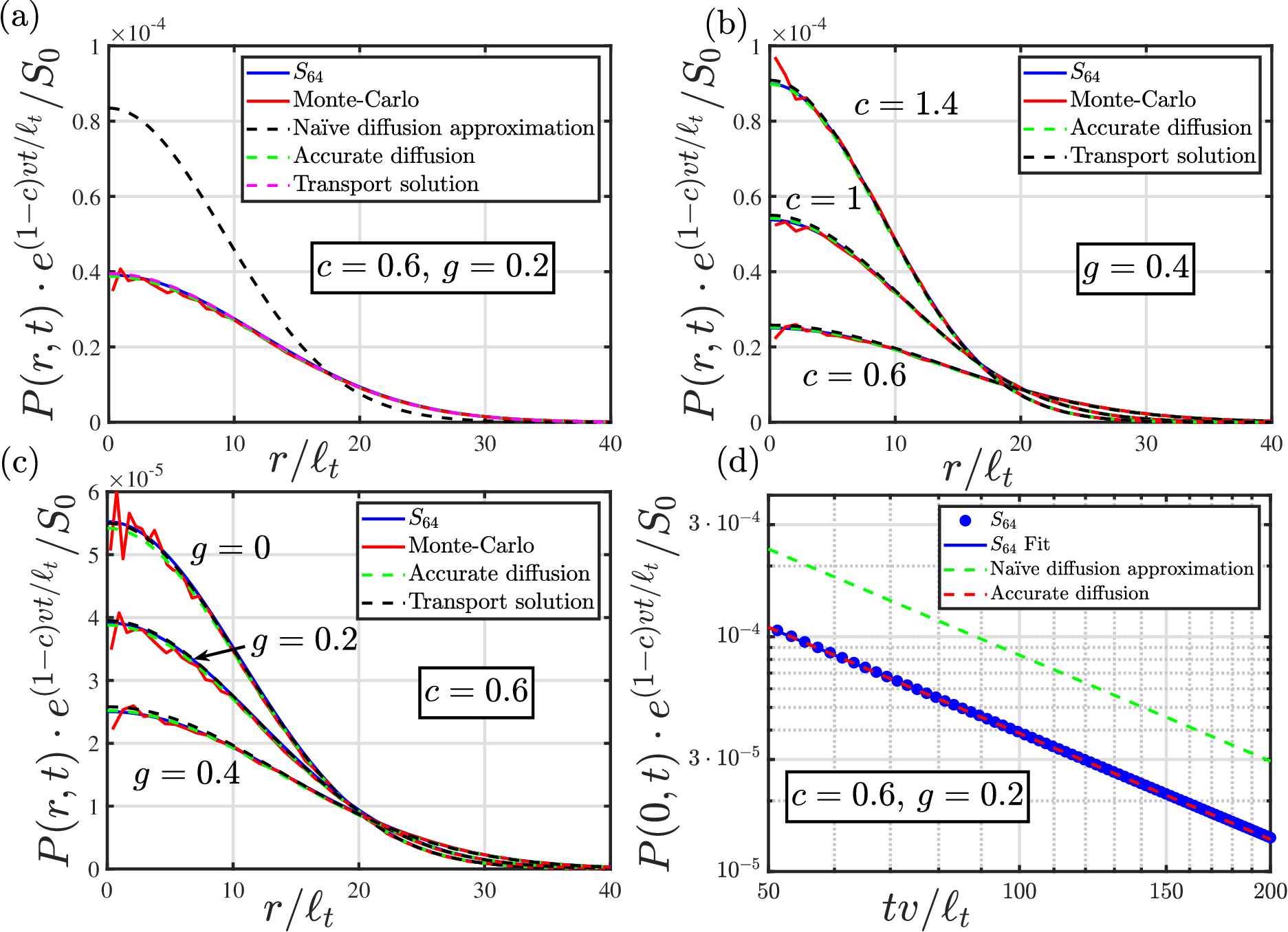}
\end{center} 
\vspace{-6.5mm}
\caption{(a) The population density ${\cal P}(r,t)$ at long times ($vt/\ell=100$), for $c=0.6$ and $g=0.2$, versus the normalized space coordinate $vt/\ell$. Shown are the full transport solutions obtained from  $S_N$ solutions (blue curves), MC simulations (red curves) and the full analytical solution~(\ref{pesach_aniso}) (magenta), which are compared with the results of the na\"ive diffusion theory (black) and the accurate diffusion coefficient~(\ref{c_diffusion}) (green curves). (b) The population density ${\cal P}(r,t)$ at long times ($vt/\ell=100$), for different values of $c$ ($g=0.2$), exact ($S_N$, MC and analytic) versus the accurate diffusion solution. (c) Same for different values of $g$ ($c=0.6$). (d) A log-log plot of the population density at the origin ${\cal P}(0,t)$, for $c=0.6$ and $g=0.2$, as a function of the normalized time $vt/\ell$. Here, results of the na\"ive  (green) and accurate (red) diffusion solutions are compared with the $S_N$ solution (blue).}
\label{long}
\vspace{-5mm}
\end{figure}

\textit{Determining the Diffusion Coefficient at Long Times.}
We now identify the effective diffusion coefficient by computing the late-time asymptotics of Eq.~(\ref{spherical}). At late times, $t\gg v/\ell_t$, the PDF is close to being isotropic and we can use the so-called {\em{diffusion assumption}}, $P(r,t,\mu)\approx {\cal P}(r,t)+3\mu {\cal P}_1(r,t)$, with ${\cal P}_1(r,t)\equiv\int_{-1}^1 \mu d\mu P(r, t, \mu)$ being the flux. Integrating Eq.~(\ref{spherical}) once over $\int\! d\mu$ and once over $\int\! \mu d\mu$ yields two equations for the zeroth and first moments of the PDF. After some algebra we obtain:
\begin{subequations}
\label{p1}
\begin{equation}
\label{p1_first}
\hspace{-2mm}\frac{\partial {\cal P}(r,t)}{v\partial t}\!+\!\frac{1}{r^2}\frac{\partial }{\partial r}\left[r^2 {\cal P}_1(r,t)\right]\!+\!\frac{1\!-\!c}{\ell_t}{\cal P}(r,t)\!=\!\frac{Q_{\mathrm{ext}}(r,t)}{v}
\end{equation}
\vspace{-4mm}
\begin{equation}
\label{p1_second}
\hspace{-3mm}{\cal P}_1(r,t)=-\frac{\ell_t}{3(1-g)}\frac{\partial {\cal P}(r,t)}{\partial r}=-\frac{D_{\mathrm{tr}}}{v}\frac{\partial {\cal P}(r,t)}{\partial r}
\end{equation}
\end{subequations}
Note that while Eq.~(\ref{p1_first}) is exact, Eq.~(\ref{p1_second}) is an approximated Fick's law, using the diffusion assumption, and neglecting the $\partial {\cal P}_1(r,t)/\partial t$ term 
compared to $(v/\ell_t){\cal P}_1(t,r)$~\footnote{Without neglecting the $\partial {\cal P}_1(r,t)/\partial t$ term, an approximated ``3d'' Telegrapher's equation is obtained.}.
Substituting Eq.~(\ref{p1_second}) in Eq.~(\ref{p1_first}) yields a of {\em{reaction-diffusion equation}}~\cite{reaction1,reaction2}, with a diffusion constant of $D_{\mathrm{tr}}\equiv v\ell_{\mathrm{tr}}/3$, where $\ell_{\mathrm{tr}}$ was defined above~\cite{case1953,CaseZweifel1967,BellGlasstone,Durduran1997,Corngold}.
The solution of Eqs.~(\ref{p1}) yields:
\begin{equation}
\label{diffusion}
{\cal P}_{\mathrm{diff}}(r,t;c)\!=\!\frac{1}{(4\pi D_{\mathrm{tr}}t)^{\nicefrac{3}{2}}}\exp\!\left[\!-\frac{r^2}{4D_{\mathrm{tr}}t}-\frac{vt(1-c)}{\ell_t}\!\right]\!.
\end{equation}
Yet, this solution yields an incorrect scaling relation ${\cal P}_{\mathrm{diff}}(r,t;c)={\cal P}_{\mathrm{diff}}(r,t;1)e^{-(1-c)vt/\ell_t}$ and thus, violates the scaling in Eq.~(\ref{scale2}) for $c\ne1$~\cite{Durduran1997,Corngold}.
This is also evident by comparing Eq.~(\ref{diffusion}) to numerical results, revealing that the approximation given by the first two moment leading to Eqs.~(\ref{p1}) breaks down at $c\ne1$ (see Fig.~\ref{long}).


To find the correct diffusion coefficient for any $c\ne1$, we use Eq.~(\ref{pesach_aniso}) in the limit of $t\to\infty$, which reads:
\begin{equation}
\label{c_diffusion}
\hspace{-2.5mm}{\cal P}(r,t\to\infty;c)\!=\!\frac{1}{\left(4\pi \frac{D_{\mathrm{tr}}}{c}t\right)^{\nicefrac{3}{2}}}\exp\!\left[\!-\frac{r^2}{4\frac{D_{\mathrm{tr}}}{c}t}\!-\!\frac{vt(1\!-\!c)}{\ell_t}\!\right]\!\!.
\end{equation}
This solution coincides with the diffusion solution~(\ref{diffusion}), but with a \textit{modified} diffusion coefficient: $D'_{\mathrm{tr}}\equiv D_{\mathrm{tr}}/c=v\ell_{\mathrm{tr}}/(3c)$, which is exact, even when the classical diffusion assumption is invalid. This result can also be obtained directly by applying the scaling relation to Eq.~(\ref{diffusion}) for $c=1$:
${\cal P}_{\mathrm{diff}}(cr,ct;c)=c^3e^{-(1-c)v/\ell_{\mathrm{tr}}}{\cal P}_{\mathrm{diff}}(cr,ct;1)$. Notably, while our result holds for any value of $c$, in the special case of $c\leqslant 1$, the diffusion constant becomes $D=v\ell_s/3$, and provides a rigorous basis for previous semi-empirical studies that obtained a similar result~\cite{Furutsu2,levpom,pomraning1981,ganapol1985,Furutsu1,Tsuchiya1995,Durduran1997,zoia}. Finally, one can also derive a diffusion equation directly from the transport equation~(\ref{spherical}) at $t\to\infty$, without explicitly assuming the diffusion assumption, 
which yields the correct diffusion coefficient in the close vicinity of $c\simeq 1$ (see SM, Sec.~V).

In Fig.~\ref{long} (a-c) we compare the probability density ${\cal P}(r,t; c)$ given by Eq.~(\ref{c_diffusion}) as function of $r$, with numerical solutions of Eq.~(\ref{spherical}) at late times, using the $S_N$ and MC methods, for different values of $g$ and $c$. The figure shows that the na\"ive choice of $D=v\ell_t/3$ yields large errors. On the other hand, using the accurate diffusion constant $D=v\ell_{\mathrm{t}}/(3c(1-g))$ yields an excellent agreement indicating that the proposed general-collision diffusion coefficient accurately reproduces the solution to the transport equation [Eq.~(\ref{pesach_aniso})] at late times, for any $c$ or $g$.

A complementary view to the spatial profiles is to use the value at the center of the Gaussian, ${\cal P}(0,t\to\infty)$ as a function of time [as can be seen from Eqs.~(\ref{diffusion}) and~(\ref{c_diffusion})], which can be numerically found by fitting the $S_N$ solution. In Fig.~\ref{long}(d) we plot ${\cal P}(0,t)e^{(1-c)vt/\ell_t}$ and compare the different diffusion coefficients from Fig.~\ref{long}(a). Again, using $D=v\ell_{\mathrm{tr}}/(3c)$ yields an excellent agreement with the $S_N$ results. However, the na\"ive choice of $D=v\ell_t/3$ yields a highly inaccurate slope.

\textit{Discussion.}
A long-standing question in the field of statistical physics, is the identification of the diffusion coefficient in the diffusion limit of the 3D transport equation, for an arbitrary set of reactions, i.e., $c\ne1$, and weakly-anisotropic scattering. 
Here we identify the correct diffusion coefficient by deriving a generalized solution for the point source Green function for the  3D spherical-symmetric transport equation. 
This was done by introducing an exact scaling relation between the general-collision ($c\ne1$) and pure-scattering ($c=1$) solutions.
We verified our analytical results using the  $S_N$  method and  Monte-Carlo simulations. 
Unlike previous results~\cite{Furutsu2,levpom,pomraning1981,Furutsu1,ganapol1985,Tsuchiya1995,Durduran1997,zoia}, the diffusion coefficient we have derived is accurate for any $c$, i.e., for arbitrary birth-death reactions. These results may provide important insight into fields such as photon diffusion in tissue optics and radiative transfer in non-equilibrium astrophysics.

\textit{Acknowledgments.}
M.A.~acknowledges support from the Israel Science Foundation Grant No.~531/20. 
\bibliography{bibliography.bib}

\onecolumngrid

\setcounter{figure}{0}
\renewcommand{\thefigure}{S\arabic{figure}}

\setcounter{table}{0}
\renewcommand{\thetable}{S\arabic{table}}

\renewcommand{\theequation}{S\arabic{equation}}    
\setcounter{equation}{0} 

\vspace{19.8cm}
\begin{center}\Huge{Supplemental Material}
\end{center}
\section{I. The Numerical Simulations}

The deterministic discrete ordinate $S_N$ simulation code was written using fully implicit scheme in time and diamond difference scheme with negative-flux-fixup in space~[36]
in {\em{Fortran}} using a constant $\Delta x/\ell_t=3\cdot 10^{-3}$ (checked for convergence), while $v\Delta t/\ell_t$ is defined dynamically such that the population density $P(r,t)$ will not change in each cell (between time steps) more than 1\%, but not greater than $v\Delta t/\ell_t=1\cdot 10^{-3}$. 

In the  Monte-Carlo simulations we have implemented a time-dependent analog Monte-Carlo method for the solution of the linear Boltzmann equation. 
Here, the particles are advanced in space and time, while scattering events are sampled from the probability distributions defined by the interaction cross sections.

\section{II. Deriving the Boltzmann Equation in Spherical Symmetry}

In a spherical geometry, $P(\vec{r}, t, \hat{\Omega})$ is a function of the position in space which is defined by the coordinates $\vec{r}=(r,\Theta,\Phi)$ ($r$ is the radial distance, $\Theta$ is the polar angle between $\vec{r}$ and the $z$-axis, and $\Phi$ is the azimuthal angle between the $x$-axis and projection of $\vec{r}$ on the $x\!-\!y$ plane), and $\hat{\Omega}=(\mu,\varphi)$ denotes the direction of the particle's velocity at any given point  (where $\mu=\cos\theta, \eta=\sin\theta\cos\varphi$ and $\xi=\sin\theta\sin\varphi$, with $\mu^2+\eta^2+\xi^2=1$, i.e. $\mu$, $\eta$ and $\xi$ are the normalized coordinates of the particle's velocity). The angular-integrated probability density (i.e., the population density) is given by $P(\vec{r},t)\equiv\frac{1}{4\pi}\int d\hat{\Omega} P(\vec{r}, t, \hat{\Omega})$. When the interaction between particles is negligible and the medium is locally in equilibrium, the Boltzmann collision term can be modeled exactly using the definitions of the mean free paths (mfps) for different interactions (or alternatively cross sections), and the equation that governs the physical process is the {\em{linear}} Boltzmann transport equation~[8, 18, 25, 36]:

\begin{equation} \label{boltzmann}
    \frac{\partial}{v\partial t}P(\vec{r}, t, \hat{\Omega}) + \hat{\Omega}\cdot\nabla P(\vec{r}, t, \hat{\Omega}) +\ell_t^{-1}P(\vec{r}, t, \hat{\Omega}) =
    \frac{1}{4\pi}\int d\hat{\Omega}' \ell_s^{-1}(\hat{\Omega}\cdot\hat{\Omega}') P(\vec{r}, t, \hat{\Omega}') + Q(\vec{r}, t, \hat{\Omega})/v.
\end{equation}
Here $v$ is the particle velocity and $Q(\vec{r}, t, \hat{\Omega})=Q_{\mathrm{int}}(\vec{r}, t, \hat{\Omega})+Q_{\mathrm{ext}}(\vec{r}, t, \hat{\Omega})$ is the sum of internal sources $Q_{\mathrm{int}}(\vec{r}, t, \hat{\Omega})$ (which depend on the moments of $P(\vec{r}, t, \hat{\Omega})$) and external sources $Q_{\mathrm{ext}}(\vec{r}, t, \hat{\Omega})$ that in general may be  functions of space, direction and time. Notably, $Q(\vec{r}, t, \hat{\Omega})$ has units of particles per unit volume per time.


In the spherically-symmetric case, the dynamics depends only on the radius $r$ and the angle $\mu=\cos(\theta)$, determining the direction of the particle with respect to $r$. As a result, $P(\vec{r}, t, \hat{\Omega})=P(r, t, \mu)$, and Eq.~(\ref{boltzmann}) takes the form of Eq.~(1) in the main text~[8, 18, 25, 36]. 

\section{III. Derivation of the scaling relation for general collision medium in 3D}
\label{ans_sec}

The Boltzmann (transport) equation in three-dimensional (3D) spherical geometry with Green function source $Q_{\mathrm{ext}}(r,t)=\delta(r)\delta(t)/4\pi r^2$ can be expressed as:
\begin{equation} \label{spherical_app}
    \left[\frac{\partial}{v\partial t} + \mu\frac{\partial}{\partial r} +\frac{1-\mu^2}{r}\frac{\partial}{\partial\mu} + \ell_t^{-1}\right]P(r, t, \mu;c) = \frac{c\ell_t^{-1}}{2}\int_{-1}^1 d\mu' f(\mu_0)P(r, t, \mu';c) + \frac{\delta(r) \delta(t)}{4\pi r^2 v},
\end{equation}
where $c(r,t)$ is given by Eq.~(2) in the main text, and is assumed to be constant.
For purely scattering media, $c=1$ and $\ell_t=\ell_s$. 


In a slab geometry, which includes an infinite planar system in the $y$ and $z$ directions and finite in $x$ direction, the  probability density function (PDF) depends only $x$ and a single angular coordinate $\theta$, the angle between $\hat{\Omega}$ and the $x$-axis. In this case, there is a known analytic scaling relation between $P(x,t,\mu;c)$ with a Green function source $Q_{\mathrm{ext}}(x,t)=\delta(x)\delta(t)$, and $P(x,t,\mu;1)$~[8,40]: 
\begin{equation}
P(x,t,\mu;c)=ce^{-(1-c)vt/\ell_t}P(cx,ct,\mu;1).
\label{scale}
\end{equation}
Similarly, we propose a scaling relation for the 3D case 
between $P(r,t,\mu;c)$ and $P(r,t,\mu;1)$ as follows:
\begin{equation}
P(r,t,\mu;c)=c^3e^{-(1-c)vt/\ell_t}P(cr,ct,\mu;1).
\label{ansatz}
\end{equation}
Substituting Eq.~(\ref{ansatz}) into~(\ref{spherical_app}) and arranging each $r$ and $t$ to have a form of $cr$ and $ct$ (including  derivatives) yields:
\begin{align}
\label{derive3}
& c^4e^{-(1-c)vt/\ell_t}\frac{\partial P(cr,ct,\mu;1)}{v\partial (ct)}+c^3e^{-(1-c)vt/\ell_t}\left[c\mu\frac{\partial}{\partial (cr)}+ c\frac{1-\mu^2}{cr}\frac{\partial}{\partial\mu} + c\ell_t^{-1}\right] P(cr,ct,\mu;1) = \\
&\frac{c^4\ell_t^{-1}e^{-(1-c)vt/\ell_t}}{2}\int_{-1}^1 d\mu' f(\mu_0)P(cr,ct,\mu';1) + \frac{c^4}{4\pi c^2r^2 v}\cdot\frac{\delta(r)}{c}\cdot\frac{\delta(t)}{c} \nonumber
\end{align}
Using the identity of $\delta(\alpha x)=1/\vert\alpha\vert\cdot\delta(x)$, the Green function source can be written as $\delta(r)\delta(t)/c^2=\delta(cr)\delta(ct)$. Thus, since $e^{-(1-c)vt/\ell_t}\to 1$ when $t\to 0$, by dividing Eq.~(\ref{derive3}) by $c^4e^{-(1-c)vt/\ell_t}$, we have:
\begin{equation}
\left[\frac{\partial}{v\partial (ct)} + \mu\frac{\partial}{\partial (cr)}+ \frac{1-\mu^2}{cr}\frac{\partial}{\partial\mu} + \ell_t^{-1}\right] P(cr,ct,\mu;1) =
\frac{\ell_t^{-1}}{2}\int_{-1}^1 d\mu' f(\mu_0)P(cr,ct,\mu';1) + \frac{\delta(cr) \delta(ct)}{4\pi (cr)^2 v}.
\label{final_ansatz}
\end{equation}
Under the transformation of $ct\to t$ and $cr\to r$, Eq.~(\ref{final_ansatz}) reproduces Eq.~(\ref{spherical_app}) with $\ell_t=\ell_s$ and $c=1$, and verifies that  Ansatz~(\ref{ansatz}) is correct.


\section{IV. The Extended Paasschens solution}

In this section we first extend the Paasschens solution for linearly anisotropic scattering cross-section, and later to an infinite homogeneous general collision media (general $c$). Although the extension to linear anisotropic scattering is well-known~[35],
we explicitly show that it converges to the exact transport solution only in the diffusion limit, i.e, for large normalized time $vt/\ell_t$ (see Fig.~3 in the main text). Yet, the extension to general $c$ is {\em{exact}} via Ansatz~(\ref{ansatz}).

\subsection{A. Extension for linearly anisotropic scattering}

The function $f(\mu_0)$ denotes the angular dependence of the scattering events, and depends on the relative cosine angles, $\mu_0\equiv\hat{\Omega}\cdot\hat{\Omega'}$. 
To account for anisotropic scattering, we expand $f(\mu_0)$ in a Legendre series~[25, 36, 41]: 
\begin{equation}
\label{leg_scat}
f(\mu_0)=\sum_{n=0}^{\infty}(2n+1)f_nP_n(\mu_0),\;\;\;\;\;\;\;f_n=\frac{1}{2}\int_{-1}^1f(\mu_0)P_n(\mu_0)d\mu_0,
\end{equation}
where $f_n$ are the scattering function moments.
Here, $f_0=1$ by definition, and $f_1=\bar{\mu}_0$.
Thus, the Boltzmann equation in this case becomes~[25, 36]: 
\begin{equation}
\label{spherical_aniso}
    \left[\frac{\partial}{v\partial t} + \mu\frac{\partial}{\partial r} +\frac{1-\mu^2}{r}\frac{\partial}{\partial\mu} + \ell_t^{-1}\right]P(r, t, \mu) = \frac{c\ell_t^{-1}}{2}\sum_{n=0}^{\infty}(2n+1)f_nP_n(\mu)\int_{-1}^1 d\mu'P(r, t, \mu')P_n(\mu') + \frac{\delta(r) \delta(t)}{4\pi r^2 v}.
\end{equation}
For  {\em{linear}} anisotropic scattering, only the first two moments are available, $f(\mu_0)=f_0+3\mu f_1$. Thus, Eq.~(\ref{spherical_aniso}) becomes:
\begin{equation}
\label{spherical_linear_aniso}
    \left[\frac{\partial}{v\partial t} + \mu\frac{\partial}{\partial r} +\frac{1-\mu^2}{r}\frac{\partial}{\partial\mu} + \ell_s^{-1}\right]P(r, t, \mu) = \ell_s^{-1}(P(r,t)+3\mu gP_1(r,t))+ \frac{\delta(r) \delta(t)}{4\pi r^2 v},
\end{equation}
where $P(r,t)\equiv\int_{-1}^1 d\mu P(r, t, \mu)$ and
$P_1(r,t)\equiv\int_{-1}^1 \mu d\mu P(r, t, \mu)$. Equation~(\ref{spherical_linear_aniso}) is exact for linear anisotropic scattering, whereas below we will assume a weak (linear) angular-dependence of the PDF. Taking the first two terms in the Legendre series, $P(r,t,\mu)\approx P(r,t)+3\mu P_1(r,t)$ (which is also used to derive diffusion approximation~\cite{BellGlasstone})[25]), and substituting this  in Eq.~(\ref{spherical_linear_aniso}) yields
Eq.~(\ref{spherical_app}) with $c=1$, and replacing $\ell_t(=\ell_s)$ with $\ell_{\mathrm{tr}}$, where $\ell_{\mathrm{tr}}^{-1}\equiv\ell_t^{-1}(1-g)=\ell_s^{-1}(1-g)$. Thus, the Paasschens-like approximate linear anisotropic scattering solution is~[35]: 
\begin{equation}
\label{pesach_aniso_app}
P(r,t)\simeq\frac{e^{-vt(1-g)/\ell_s}}{4\pi r^2}\delta(r-vt)+\frac{(1-r^2/v^2t^2)^{\nicefrac{1}{8}}}{(4\pi\ell_s vt/3(1-g))^{\nicefrac{3}{2}}}e^{-vt(1-g)/\ell_s}\, G\left(\frac{vt(1-g)}{\ell_s}\left[1-\frac{r^2}{v^2t^2}\right]^{\nicefrac{3}{4}}\right) \Theta(vt-r).
\end{equation}

\subsection{B. Extension for general media}

We now extend the Paasschens solution for general $c$. To do so, we combine the extended solution that includes anisotropic scattering [Eq.~(\ref{pesach_aniso_app})],  derived for purely scattering media (i.e. $c=1$), with the exact Ansatz~(\ref{ansatz}). Employing the identities: $\delta(\alpha x)=1/\vert\alpha\vert\cdot\delta(x)$ and $\Theta(\alpha (x-\beta))=\Theta(x-\beta)\Theta(\alpha)+\Theta(-(x-\beta))\Theta(-\alpha)$, the  extended-Paasschens  solution for general $c$ becomes:

\begin{equation}
\label{pesach_sofi}
P(r,t;c)\simeq\frac{e^{-vt(1-cg)/\ell_t}}{4\pi r^2}\delta(r-vt)+\frac{(1-r^2/v^2t^2)^{\nicefrac{1}{8}}}{(4\pi\ell_tvt/[3c(1-g]))^{\nicefrac{3}{2}}}e^{-vt(1-cg)/\ell_t}\, G\left(\frac{vct(1-g)}{\ell_t}\left[1-\frac{r^2}{v^2t^2}\right]^{\nicefrac{3}{4}}\right) \Theta(vt-r)
\end{equation}


\section{V. From Boltzmann transport equation to diffusion at long times}
\label{TD_ficks}

Here, we derive a diffusion equation directly from the  Boltzmann equation in a uniform medium at $t\to\infty$, without explicitly assuming an isotropic PDF (the diffusion assumption). The derivation is done in a 1D slab geometry, but the extension to 3D is straightforward, via the Peierls integral transport equation~[24]. 

The time-dependent one-velocity isotropic-scattering Boltzmann equation in a slab-geometry is:
\begin{equation}
\frac{1}{v}\frac{\partial P(x, t, \mu)}{\partial t}+\mu\frac{\partial
P(x, t, \mu)}{\partial x}+\ell_t^{-1}P(x,t,\mu)=
\frac{c\ell_t^{-1}}{2}\int_{-1}^1{P(x, t, \mu')d\mu'}.
\label{mono_time}
\end{equation}
Performing a Laplace-transform in the time coordinate, we arrive at: 
\begin{equation}
\mu\frac{\partial \hat{P}_s(x,\mu)}{\partial x}+\hat{\ell}_{t,s}^{-1}
\hat{P}_s(x,\mu)=\frac{\hat{c}_s\hat{\ell}_{t,s}^{-1}}{2}\int_{-1}^1{
\hat{P}_s(x,\mu')d\mu'},\quad\quad
\hat{\ell}_{t,s}^{-1}\equiv\ell_t^{-1}+\frac{s}{v},\quad\quad \hat{c}_s=\frac{c}{1+\frac{s\ell_t}{v}},
\label{new_albedo_sigma}
\end{equation}
where we have defined a modified $s$-dependent total cross-section $\hat{\ell}_{t,s}^{-1}$ and  mean number of particles that  are emitted per collision, $\hat{c}_s$t~[24]. 
This equation has the form of the time-independent Boltzmann
equation, which can be solved using well-known techniques~[8, 24]. 
Assuming a separation of variables of 
$\hat{P}_s(x,\mu)=e^{-\hat{\kappa}_s\hat{\ell}_{t,s}^{-1}x}f(\mu)$,
with $\hat{\kappa}_s$ as the $s$-dependent eigenvalues: 
$2/\hat{c}_s=\hat{\kappa}_s^{-1}\ln\left[(1+\hat{\kappa}_s)/(1-\hat{\kappa}_s)\right]$,
the time-dependent asymptotic $\hat{P}_s(x,\mu)$ 
becomes:
\begin{equation}
\hat{P}_s(x,\mu)=\frac{\hat{c}_s}{2}A_0\frac{e^{\hat{\kappa}_{s,0}
\hat{\ell}_{t,s}^{-1}x}}{1+\mu\hat{\kappa}_{s,0}}+\frac{\hat{c}_s}{2}B_0
\frac{e^{-\hat{\kappa}_{s,0}\hat{\ell}_{t,s}^{-1}x}}{1-\mu\hat{\kappa}_{s,0}}.
\label{asymptotic_flux}
\end{equation}
The zeroth and first moments of the asymptotic $\hat{P}_s(x,\mu)$ satisfy
a modified Fick's law of this form: $
\hat{P}_{1,s}(x)=-\hat{D}_s(\hat{c}_s,\hat{\ell}_{t,s})\cdot\partial
\hat{P}_s(x)/\partial x$,
with a $s$-dependent modified diffusion coefficient:
\begin{equation}
\hat{D}_s(\hat{c}_s,\hat{\ell}_{t,s}^{-1})\equiv\frac{1-\hat{c}_s}{
\hat{\kappa}_{s,0}^2(\hat{c}_s)\hat{\ell}_{t,s}^{-1}}=v\frac{v\ell_t^{-1}(1-c)+s}{(v\ell_t^{-1}+s)^2
\hat{\kappa}_{s,0}^2(\hat{c}_s)}\equiv v\frac{D_0(\hat{c}_s)}{v\ell_t^{-1}+s},
\label{new_d_s1}
\end{equation}
where  
$D_0(\hat{c}_s)$ satisfies a transcendental equation~[11, 14, 23, 24]. 
Note that, the asymptotic solution, Eq.~(\ref{asymptotic_flux}), is exact {\em{only}} for $c=1$, while for $c\ne 1$, the ratio of the transient part 
increases.
Therefore,  we focus on the limit $\vert1-c\vert\ll 1$ at long times, i.e.,  $s\to 0$. Indeed, expanding Eq.~(\ref{new_d_s1})  around $s=0$,
one obtains Fick's law 
in limit of $\vert1-c\vert\ll 1$, where $\hat{c}_s\to c$:
$P_1(x,t)=-D_0(c)\ell_t\cdot\partial P(x,t)/\partial x$.

Since we are interested in the limit of $\hat{c}_s\to c$, we present the dependency of $D_0(c)$ near $c=1$~[42] 
\begin{equation}
D_0(c)=
\frac{4}{\pi^2c}\left(\frac{c-0.0854}{c+0.112}\right)\approx\frac{1}{3c}\left[1+0.2(1-c)\right].
\label{d_win}
\end{equation}
Note that, taking the limit of $\vert1-c\vert\ll 1$, which is the regime where Eq.~(\ref{asymptotic_flux}) is exact, one obtains in the leading order a diffusion coefficient of $1/(3c)$, instead of $1/3$. In addition, in~[43] 
there is a simple but elegant approximate expression for $\kappa_0(c)$ which is an extrapolation from the 1D and 2D exact Boltzmann solutions, to the 3D case:
\begin{equation}
D_0(c)=\frac{(1-c)}{\kappa_0^2(c)}=\frac{(1-c)}{1-c^3}=\frac{1}{1+c+c^2}\approx\frac{1}{3c}\left(1-\frac{1}{3}(1-c)^2\right).
\label{eugene_D}
\end{equation}
This result coincides with our result in the main text, $1/(3c)$, for the general collision scenario, up to  second order. 

\end{document}